\documentclass[preprint,aps,floatfix,showpacs]{revtex4}
\usepackage{graphicx}
\def\bea{\begin{eqnarray}}
\def\eea{\end{eqnarray}}
\begin{document}
\preprint{JLAB-THY-02-35}

\title{Negative Parity $\mathbf{70}$-plet Baryon Masses  in the  $\mathbf{1/N_c}$ Expansion}

\author{
J. L. Goity $^{a,b}$ \thanks{e-mail: goity@jlab.org}, \
C. L. Schat $^{a,b}$ \footnote[2]{Present address: Department of Physics, Duke University, Durham, NC 27708} 
 $^\ddag$ \thanks{e-mail: schat@jlab.org}, \
N. N. Scoccola $^{c,d}$ 
 \footnote[3]{Fellow of CONICET, Argentina.}
\thanks{e-mail: scoccola@tandar.cnea.gov.ar}}

\affiliation{
$^a$ Thomas Jefferson National Accelerator Facility, Newport News, VA 23606, USA. \\
$^b$ Department of Physics, Hampton University, Hampton, VA 23668, USA. \\
$^c$ Physics Department, Comisi\'on Nacional de Energ\'{\i}a At\'omica,
     Av.~Libertador 8250, (1429) Buenos Aires, Argentina.\\
$^d$ Universidad Favaloro, Sol{\'\i}s 453, (1078) Buenos Aires, Argentina.}

\date{\today}
\begin{abstract}

The masses of the negative parity $SU(6)$ {\bf 70}-plet baryons
are analyzed in the $1/N_c$ expansion to order $1/N_c$ and to
first order in $SU(3)$ breaking. At this level of precision there
are twenty predictions. Among them there are the well known
Gell-Mann Okubo and equal spacing relations, and four new
relations involving $SU(3)$ breaking splittings in  different
$SU(3)$ multiplets. Although the breaking of  $SU(6)$ symmetry
occurs at zeroth order in $1/N_c$, it turns out to be small. The
dominant source of the breaking is the hyperfine interaction which
is of order $1/N_c$. The spin-orbit interaction,  of zeroth order
in $1/N_c$, is entirely fixed by the splitting between the singlet
states $\Lambda(1405)$ and $\Lambda(1520)$, and the spin-orbit
puzzle is solved by the presence of other zeroth order operators
involving flavor exchange.

\end{abstract}

\maketitle

\section{Introduction}

The understanding of baryons and their excitations from QCD still represents  a wide open chapter in strong
interaction physics. This is naturally so as baryons are the sector of the strong interactions where the
non-perturbative QCD dynamics is likely to be  most difficult to sort out. With some exceptions, the current
understanding of the baryon sector is largely based on data of different sorts and its analyses by means of models,
most prominently the constituent quark model in its different  versions \cite{CandR}. One exception are the ground
state baryons, namely the spin 1/2 octet and spin 3/2 decouplet, where definite progress in relating observables to
QCD has been achieved by means of effective theories. At low energies Chiral Perturbation Theory (ChPT) is such an
effective theory\cite{ChPT}. In addition, the implementation of the $1/N_c$ expansion\cite{RevMan,RevJen,RevLeb},
where $N_c$ is the number of colors in QCD, brings in constraints on the effective couplings entering in ChPT
\cite{Flores} increasing in this way the predictive power of the effective theory. Although this effective theory
has validity within a limited low energy domain, it represents QCD faithfully, and therefore its predictions are
genuine QCD predictions.

Beyond the low energy domain the availability of effective theories has been  limited. In particular, in the
resonance domain (excited baryons) there has not been a well established model independent analysis scheme. In order to formulate an effective theory, it is necessary to identify the small expansion parameters available.
In the resonance domain, besides the light quark masses, one such expansion parameter is provided by $1/N_c$. The  expansion in $1/N_c$  for excited baryons has been proposed and applied in several works 
\cite{CGO94,Goi97,PY98,CCGL,SGS02,Schat,Azi01} with encouraging success.

The importance of having a model independent approach to the physics of excited baryons should be emphasized. On
one hand it is still mysterious how and why  constituent quark models are  a good qualitative, and at times also
quantitative, description of that sector. It is not quite clear either what the specific deficiencies  of the
different versions of the quark model are. On the other hand, there is currently important experimental improvement
being achieved in  particular thanks to the $N^*$ program at Jefferson Lab \cite{CLAS}, where the
study of  non-strange resonances is being carried out with unprecedented precision, and also there is important
progress towards studying the baryon spectrum by means of lattice QCD simulations \cite{Ric01}. This promising developments have rekindled the theoretical interest in excited baryons \cite{Simonov}.  Sorting out and
understanding the physics emerging from the old and new experimental data as well as from the lattice would be
greatly optimized if an effective theory is available.

Based on general arguments,  it is believed that an expansion in $1/N_c$ as first proposed by 't~Hooft \cite{tHo74}
should hold in QCD in all regimes. The application of the $1/N_c$ expansion to baryons starting with  the
pioneering work  of Witten \cite{Witten} has served to understand numerous issues. In the past several years the
$1/N_c$ expansion in the ground state baryons has been extensively investigated as summarized in several reviews
\cite{RevMan,RevJen,RevLeb}.

Although baryons  in the large $N_c$ limit  belong to  increasingly large representations of the spin and flavor
groups, thus giving the appearance that they are  increasingly complicated, there is instead a great degree of
simplification emerging in that limit. In the ground state sector  it was discovered
\cite{GeSa84,Dashen1} that the requirement that unitarity be fulfilled in $\pi$-$N$ elastic scattering in the large $N_c$ limit requires a dynamical spin-flavor symmetry that leads to the mentioned simplification. The reason
unitarity gives rise to this non-trivial symmetry constraint is that  the pion-nucleon coupling scales as
$\sqrt{N_c}$, leading naively to a scattering amplitude of order $N_c$, which for large $N_c$  would imply  a
violation of unitarity. The dynamical symmetry is a  contracted $SU(2F)$  spin-flavor
symmetry\cite{GeSa84,Dashen1}, where $F$ is the number of flavors. Up to corrections of order $1/N_c$ it is
possible to replace the contracted symmetry group by the ordinary $SU(2F)$ group \cite{LM94}.  The possibility of
building an effective theory based on this dynamical symmetry rests on the fact that for ground state baryons the
symmetry is broken  at  order $1/N_c$. It is therefore possible to implement the $1/N_c$ expansion around the
spin-flavor symmetry limit. Once the spin-flavor multiplet has been identified, which in the case of ground state
baryons is the totally symmetric $SU(2F)$ representation with $N_c$ fundamental indexes, 
the $1/N_c$ expansion of different observables can be carried out in terms of an expansion in composite operators
\cite{Dashen1,Jenk1,Jenk2,DJM2,JL95} which are sorted according to their order in $1/N_c$. For different static
observables, such as masses, magnetic moments, axial couplings, etc., using  the Wigner-Eckart theorem a basis of
operators can be  built in terms of products of the generators of the  spin-flavor group. This method was applied
in the sixties \cite{Greenberg67}, and when combined with the $1/N_c$ expansion it has lead to very successful
analyses of the ground state baryon masses \cite{Dashen1,Jenk1,Jenk2,Jenk3,DJM2,JL95}, magnetic moments \cite{Jenk3,Dai}, quadrupole moments \cite{Buchmann},
and axial couplings \cite{Dai}.

While the $1/N_c$ expansion is implemented  rigorously along those lines for the ground state baryons, for excited
baryons there is one difficulty of principle. This has to do with
the observation \cite{Goi97}, to be made more explicit later, that
in general spin-flavor symmetry is not exact for excited baryons
even in the $N_c\to \infty$ limit. 
The breaking of spin-flavor symmetry at zeroth order in $1/N_c$ is  identified in the
constituent quark picture by the coupling of the orbital angular momentum. Such a  breaking
can give rise to spin-flavor configuration mixing, {\it i.e.} mixing of different spin-flavor representations, at zeroth order. As shown elsewhere \cite{GS}, configuration mixing is also driven by the coupling of orbital angular momentum. This would suggest that spin-flavor symmetry cannot be used to formulate the
$1/N_c$ expansion. However, it is well established from phenomenology that the orbital angular momentum couples very weakly. This is shown in analyses in the quark model  \cite{IK1,IC} as well as analyses in the $1/N_c$ expansion along the lines followed in this work. The  zeroth order  spin-flavor symmetry breaking turns out to be in the real world with $N_c=3$ similar in magnitude to that of order $1/N_c$ breaking effects. This is illustrated by comparing the spin-orbit splitting of $115~{\rm MeV}$ between the negative parity singlet $\Lambda$s, namely the
$\Lambda(1405)$ and the $\Lambda(1520)$, to the hyperfine splittings that are of order $1/N_c$ and about
$150~{\rm MeV}$ in the corresponding negative parity states. This suggests that configuration mixing will as well be small. With this latter assumption, it is appropriate in practice to neglect configuration mixing in studying the spectrum to a level of precision of order $1/N_c$ when $N_c=3$. In the present analysis of  the negative parity baryons that belong primarily to the  {\bf 70}-plet of $SU(6)$, a systematic error of order  $\delta_{\rm mix}^2/\Delta M$ where $\delta_{\rm mix}$ is the mixing component of the mass Hamiltonian and 
$\Delta M$ is the splitting between the {\bf 70}-plet and the
 {\bf 56}-plet with which it mixes.  
Unfortunately no solid information exists about negative parity baryons that could be assigned primarily to a {\bf 56}-plet, and therefore no convincing estimate can at present be made about that systematic error. 
Thus, with the hypothesis that configuration mixing is small, the implementation of the $1/N_c$ expansion for excited baryons masses  by working within a spin-flavor multiplet  can be carried out along similar lines as in the ground state baryons as it was shown in \cite{Goi97,CCGL,SGS02}.

It should be noticed that excited baryons are not narrow in the large $N_c$ limit \cite{CGO94,PY98,CC00} as the coupling of
pions and kaons mediating the transitions to ground state baryons are of order $N_c^0$. For this reason, the
possibility that excited baryons are built as meson-baryon resonances in large $N_c$ is quite open. Indeed, such a
possibility is well illustrated in the Skyrme model where excited baryons are built as resonances in
meson-baryon scattering \cite{Mattis}. Resonance models are being currently studied by various
groups \cite{Lee02}. At this point it is important to emphasize that the $1/N_c$ analysis does not imply a specific
picture of the excited baryons, as it relies entirely on group theoretical arguments and in ordering effective
operators in powers of  $1/N_c$, and will in particular include the possibility that excited baryons are to a large extent resonances.

In the early study of negative parity baryon masses the non-strange baryons were considered, where  the expansion was
carried out up to order $1/N_c^2$ \cite{CCGL}. Later the extension to  states with non-vanishing strangeness was
made in Ref. \cite{SGS02} where the $SU(6)$ {\bf 70}-plet masses were  analyzed  to order $1/N_c$ and to order $\epsilon$ where this latter
parameter is a measure of the magnitude of $SU(3)$ breaking by the strange quark mass. In this work that analysis is presented in detail. 

This paper is organized as follows: section II presents the construction of the negative parity baryon states,
section III gives the basis of operators and their matrix elements in the {\bf 70}-plet states, section IV presents
the fit to the known {\bf 70}-plet masses and mixings together with a discussion of the results of the fit, and finally
section V gives a general discussion and conclusions.

\section{Negative Parity Baryon States}

The excited baryon states that correspond in the constituent quark model to the first radial and orbital
excitations fit quite well into respectively  a positive parity ${\bf 56}^+$ and a negative parity ${\bf 70}^-$ irreducible representation of the spin-flavor group $SU(6)$. Although in both cases not all the states have been
experimentally determined, it seems that with those that are  well established, namely those assigned a status of at least
three stars by the Particle Data Group \cite{PDG}, it is safe to establish that observation. This conclusion is reinforced by the success of the analyses of the masses in both groups of states \cite{CCGL,SGS02,CC00}. 

In a constituent quark picture and in the large $N_c$ limit, the lowest baryonic excitations consist of a core of
$N_c-1$ quarks in the ground state of the Hartree effective potential, the core being therefore in the totally
symmetric spin-flavor representation, and an excited quark, which for the negative parity baryons discussed in this
work is in an $\ell=1$ state \cite{Goi97,CGO94}. The states therefore fill the $(3,{\bf 70})$ multiplet of the group $O(3)\otimes SU(6)$. Phenomenology strongly indicates that even when the current quark masses are small and the constituent quark picture is most likely not accurate, the 
states can still be identified as belonging to the  $(3,{\bf 70})$. Since for  the group theoretical aspects of the analysis in this work  the use of the constituent quark picture can be made with  no loss of generality, throughout the constituent quark terminology will be often used.

In order to explicitly build the states, it is first convenient to give a brief review of the $SU(6)$ group. It
has thirty five  generators, namely  $\{S_i,T_a,G_{ia}\}$, with $i=1,2,3$ and $a=1,\cdots,8$,  where the first
three are the generators of the spin $SU(2)$, the second eight are the generators of flavor $SU(3)$, and the last
twenty four can be identified as an octet of axial-vector  currents in the limit of zero momentum transfer. The
algebra of $SU(6)$ has the following commutation relations that fix the normalizations of the generators:

\bea
\left[S_i,S_j\right] &=& i\epsilon_{ijk} S_k \nonumber \\
\left[T_a,T_b\right] & =& i f_{abc} T_c \nonumber \\
\left[G_{ia},G_{jb}\right] & = & \frac{i}{4} \delta_{ij} f_{abc} T_c+
\frac{i}{2} \epsilon_{ijk}(\frac{1}{3}\delta_{ab} S_k+d_{abc} G_{kc}) \nonumber \\
\left[S_i,G_{ja}\right] & = & i\epsilon_{ijk}G_{ka} \nonumber \\
\left[T_a,G_{ib}\right] & = & if_{abc}G_{ic},
\eea
where $d_{abc}$ and $f_{abc}$ are the usual $SU(3)$ symmetric and antisymmetric  tensors, respectively.
In the non-relativistic quark picture these generators can be expressed in terms of the quark fields:
\bea
S_i=q^{\dagger} \frac{\sigma_i}{2} q  \qquad , \qquad
T_a=q^{\dagger} \frac{\lambda_a}{2} q \qquad , \qquad
G_{ia}=q^{\dagger} \frac{\sigma_i\lambda_a}{4} q,
\eea
where the Gell-Mann matrices are normalized as ${\rm Tr} \lambda_a \lambda_b=2\delta_{ab}$.

The states in the totally symmetric irreducible representation S are given by a Young tableau that consists of a
single row of  $N_c$ boxes, and the mixed symmetric irreducible representation MS relevant to this work consists of
a row with $N_c-1$ boxes and a second row with one box.

In order to build the states belonging to the mixed symmetric $SU(6)$ irreducible representation
it is convenient to start by considering the states
\bea \label{wfsc}
 |S\, S_z ; (p,q),Y, I\, I_z  ;  S^c>  &= & \sum
\left(
    \begin{array}{cc|c}
        S^c   & \frac{1}{2} & S   \\
        S^c_z & s_z         & S_z
    \end{array}
\right)
\left(
    \begin{array}{cc|c}
(p^{c} , q^{c})  &  (1 , 0)     &  (p,q)  \\
(Y^c, I^c\, I_z^c)  &  (y ,  \frac{1}{2}  \, i_z ) &  (Y, I\, I_z)
    \end{array}
\right)   \\
&\times &
\mid S^{c}\, S^{c}_z ; (p^{c} , q^{c}),   Y^c, I^c\, I_z^c\rangle
\mid \frac{1}{2}\,  s_z;  (1 , 0),  y , \frac{1}{2}  \, i_z\rangle \nonumber  ,
\eea

where $S$ is the the total spin of the baryon associated with the spin group $SU(2)$, $S^c$ is the core spin
(the core is in the S representation of SU(6)), $Y$ and $I$ are the hypercharge and isospin
respectively, and $(p,q)$ indicates the  $SU(3)$ irreducible representation. For $SU(3)$ a Young tableau  denoted by the pair
$(p,q)$ consists of $p+q$ boxes in the first row and $q$ boxes in the second.
 From the decomposition of the S  representation of $SU(6)$
as a sum of direct products of  irreducible representations of $SU(2)\otimes SU(3)$ it results that $p^c+2 q^c=N_c-1$
and $p^c=2 S^c$. This latter relation is a consequence of the fact that  for the S representation the two factors in the direct
products involved in the decomposition  have the same Young tableau $(p,q)$. The rule then results from the fact
that $p=2 S$ in  $SU(2)$.   The   $p^c=2 S^c$ relation is a generalization of the so called $I=J$ rule for two
flavors. The $SU(2)$ Clebsch-Gordan coefficients are defined with the standard Condon-Shortley phase convention
(see e.g. Ref.\cite{Edm57}) while for the extra phase conventions needed to completely specify the $SU(3)$
Clebsch-Gordan coefficients the conventions in  Ref. \cite{Hec62,Ver68} are followed.

Not all the  states displayed in Eqn. (\ref{wfsc}) are in the MS irreducible representation of $SU(6)$.
While states with $p\neq 2 S$  belong automatically in the MS representation, those with $p = 2 S$
are a linear combination of states in the S and MS representations.  The corresponding S and MS states are easily
obtained  by considering in each representation the quadratic Casimir invariant of SU(6), namely
$C^{(2)}_{SU(6)} = 2 \ G_{ia} G_{ia} + \frac{1}{2} C_{SU(3)}^{(2)} + \frac{1}{3} C_{SU(2)}^{(2)}$.
For the S  representation 
$C^{(2)}_{SU(6)}=5N_c(N_c+6)/12$, and  for the MS  representation  
$C^{(2)}_{SU(6)}=N_c(5N_c+18)/12$. Making use of these relations  the ${p}=2 S$ states in the MS  representation turn out to be  given by
\bea
\mid S\, S_z; (p=2 S,q), Y, I\,  I_z>\!\!\!_{_{\rm MS}}\!\!\! & = &\!\!\!
\sqrt{\frac{S(N_c+2(S+1))}{N_c(2S+1)}}\mid S\, S_z; (p,q), Y, I\, I_z;\ S^c=S+\frac{1}{2}> \nonumber \\-\!\!\! &&\!\!\!\!\!
\sqrt{\frac{(S+1)(N_c-2S)}{N_c(2S+1)}}\mid S\, S_z; (p,q), Y, I\, I_z;\ S^c=S-\frac{1}{2}>\! .
\eea
The states belonging to the (3,{\bf 70}) of $O(3)\otimes SU(6)$ are now expressed by including the orbital angular momentum:
\begin{equation}
 |J\, J_z  ;  S; (p,q), Y, I\, I_z >\!\!\!_{_{\rm MS}}
= \sum
\left(
    \begin{array}{cc|c}
        S   & \ell & J   \\
        S_z & m & J_z
    \end{array}
\right)
 |S\, S_z  ; (p,q), Y,I\,I_z >\!\!\!_{_{\rm MS}}\mid \ell\  m \rangle,
\end{equation}
where $J$ is the total angular momentum of the baryon.
For $N_c=3$ the negative parity states span the (3,{\bf 70}) irreducible representation. Expressing them in
the obvious notation $^{2S+1}d_J$, they are as follows: five $SU(3)$ octets ($^28 _{\frac{1}{2}}$, $^28 _{\frac{3}{2}}$,
$^48 _{\frac{1}{2}}$, $^48 _{\frac{3}{2}}$, $^48 _{\frac{5}{2}}$), two decouplets
($^2 10_{\frac{1}{2}}$, $^2 10_{\frac{3}{2}}$), and two singlets
($^2 1_{\frac{1}{2}}$, $^2 1_{\frac{3}{2}}$). The corresponding states in these
multiplets are shown in the first column of Table \ref{tab6}.

The octets $(p=1)$ of spin 1/2 satisfy $p=2 S$ and therefore involve a linear combination of core states as specified in
Eqn. (4). All other states have $p\neq 2 S$ and have therefore  core states with well defined spin: the spin 3/2
octets as well as the decouplets ($p=3$) have $S^c=1$, and the two singlet $\Lambda$ states ($p=0$) have $S^c=0$.

In the limit of exact $SU(3)$ symmetry there are two possible mixings induced by interactions that break
spin-flavor symmetry. These mixings are between the pairs of states that are in the two octets with same $J$. The
mixing angles are defined according to:
\bea
\left(
    \begin{array}{c}
        8_{J}  \\
        8'_{J}
    \end{array}
\right)
&=&
\left(
    \begin{array}{rr}
        \cos{\theta_{2 J }}  & \sin{\theta_{2 J }}  \\
            - \sin{\theta_{2 J }}  &  \cos{\theta_{2 J }}
    \end{array}
\right)
\left(
    \begin{array}{c}
        ^28_{J}  \\
        ^48_{J}
    \end{array}
\right) \ ,
\eea
where $J=\frac{1}{2}$ and $\frac{3}{2}$, $8^{(')}_J$ are mass eigenstates, and
the mixing  angles are constrained to be in the interval $[0 , \pi)$.
At this point it is easy to check that in the $SU(3)$ limit there are nine masses and
two mixing angles, i.e., eleven observables.

\section{Mass Operators}

In the subspace of the MS states, the mass operator can be expressed in terms of a linear combination of composite
operators sorted according to their order in $1/N_c$. A basis of such composite operators can be constructed using
the $O(3)\otimes SU(6)$ generators, distinguishing two sets according to whether the generator acts on the core or
on the excited quark.  Generators acting on the core will be denoted by $\{S^c_i, T^c_a, G^c_{ia}\}$ and those
acting on the excited quark by $\{s_i,t_a,g_{ia}\}$. Operators can be classified according to their n-body
character. Thus, operators containing a product  of  $n$ $SU(6)$ generators acting only  on the core, and operators
containing  a product of $n-1$ $SU(6)$ generators acting on the core and at least one generator of
$O(3)\otimes  SU(6)$
acting on the excited quark, are said to be $n$-body operators. The $1/N_c$ counting can then be determined
by the following two criteria:

i) An $n$-body operator requires that at least $(n-1)$ gluons be  exchanged between the $n$ quarks, and thus, the
coefficient that multiplies the operator in the effective theory is proportional to $1/N_c^{(n-1)}$. Henceforth,
this factor will be absorbed in the definition of the operator.

ii) In the large $N_c$ limit  the   $SU(6)$ generators  $G^c_{ia}$ with $a=1,2,3$, and $T_8^c$ have matrix elements of order $N_c$ between states with spin and strangeness of order $N_c^0$, and they are therefore called coherent generators.  The presence of coherent factors in a
 composite operator  leads to an enhancement of its matrix elements between such states
 given by a power of $N_c$. In order to determine that power it is necessary to first reduce the products of generators  
by means of the commutation relations. 

At this point it should be noted that there is an ambiguity in the identification of the physical 
states for $N_c=3$ with the states in the large $N_c$ spin-flavor multiplet. 
As in previous works, we resolve this ambiguity by identifying the physical 
states with those of spin and strangeness of order $N_c^0$ so that (ii) 
can be applied.  

In the construction of composite operators  identities for
certain products of generators valid in a  given irreducible
representation of $SU(6)$ should be used.  These identities or reduction rules
have been given  in Ref.
\cite{DJM2} for the case of the S
representation of $SU(6)$. Those  relevant to  the present work are applied to products
of generators acting on the core and are the following ones:
\bea
12\{G^c_{ia}, G^c_{ia}\} &=&
5 (N_c - 1) (N_c+5)-4 S_c^2-3\{T^c_a,T^c_a\}
\nonumber\\
 \{T^c_a,T^c_a\} &=&
2 S_c^2+\frac{(N_c - 1) (N_c + 5)}{6} \nonumber  \\ 
\{G^c_{ia},G^c_{ja}\}\mid_{J=2} &=& \frac{1}{3}
\{S^c_i,S^c_j\}\mid_{J=2} \nonumber\\
d_{abe}\{G^c_{ia},G^c_{jb}\}\mid_{J=2} &=&
\frac{1}{3}\{S^c_i,G^c_{je}\}\mid_{J=2}. \eea In particular, the
first reduction rule  results from the quadratic
Casimir, the second results from the $p^c=2 S^c$ rule mentioned
earlier.

The mass operators in the basis must be rotationally invariant, parity  and time reversal even
and isospin symmetric. At order $N_c$ the basis of $SU(3)$ singlet operators
consist of one  operator, namely the identity operator that essentially
counts the number of valence quarks in the baryon and, therefore, preserves
spin-flavor symmetry. At order $N_c^0$ only operators involving factors of
the orbital angular momentum appear.
It is not difficult to understand why
there are operators of  order $N_c^0$:  by looking at the constituent quark
picture, the excited quark moves in the effective potential of  order
$N_c^0$ generated by the core giving  rise to a spin-orbit interaction
(e.g. the $\ell \cdot s$ operator)  with a strength of order $N_c^0$.
The matrix
elements of the excited quark spin in the MS representation 
are of order $N_c^0$, and so the spin-orbit interaction is of that
order as well
\cite{footone}.
There are three
linearly independent operators of order $N_c^0$, the operators $O_2$, $O_3$
and $O_4$ listed in Table \ref{tab1}. The 1-body operator $O_2$ is the ordinary spin-orbit
operator, while the remaining operators are 2-body and involve factors
carrying flavor. The dynamics giving rise to composite operators involving
flavor exchange is not understood but it is likely that  long distance
effects due to the pion and kaon clouds give  an important part of the
strength to these operators. In particular   pion-exchange quark models
\cite{Glozman,Collins} lead naturally to important flavor exchange
contributions. It is a straightforward exercise to check that any other
operators  of order $N_c^0$ are   linearly dependent with the identity and the
three  operators of order $N_c^0$  up to terms of order $1/N_c$.
There are
eleven operators  of  order $1/N_c$  in the basis. Since for $N_c=3$ there
are eleven $SU(3)$ singlet mass observables as indicated earlier, and four
basis operators have already been identified at orders $N_c$ and $N_c^0$,
it results that for $N_c=3$ only seven operators of order $1/N_c$ are independent.
We consider those listed in Table \ref{tab1}. Among them
is the very important hyperfine operator $O_6$, known to play a crucial role
in baryon spectroscopy. Note that there are also three 3-body operators as
well.
At this point it is opportune to give a further understanding of the
singlet operators based on group theory. For $N_c=3$ the $SU(3)$ singlets
are those contained  in the $SU(2)\otimes SU(3)$ decomposition of the product
$\overline{{\bf 70}} \otimes {\bf 70}$, namely:
\bea
\overline{{\bf 70}} \otimes {\bf 70} & = &
4 (1,{\bf 1}) \oplus
5 (3,{\bf 1}) \oplus
2 (5,{\bf 1}) \oplus (7,{\bf 1}) \nonumber \\
& &  \oplus
7 (1,{\bf 8}) \oplus
11(3,{\bf 8}) \oplus
6 (5,{\bf 8}) \oplus
2 (7,{\bf 8}) \oplus  ...
\eea
which shows that there are  four  singlet operators  with $\ell = 0$  ($O_1,~O_6,~O_7,~O_{10}$), five  with  $\ell = 1$
($O_2,~O_4,~O_5,~O_9,~O_{11}$),
two  with   $\ell = 2$ ($O_3,~O_8$) and one operator with  $\ell = 3$  that
obviously cannot contribute to matrix elements between states with $\ell=1$.

The breaking of $SU(3)$ symmetry is driven by the mass difference $m_s-\hat{m}$, where $\hat{m}$ is
the average of the u- and d-quark masses.  A measure of
the breaking is given by the ratio $\epsilon=(M_K^2-M_\pi^2)/\Lambda^2$
where $\Lambda$ is a light hadronic scale, for instance a vector meson mass.
Here $SU(3)$ breaking will be included  to order $\epsilon$.
Note that for $N_c=3$, $\epsilon$ and $1/N_c$ are of similar size, and therefore  corrections of order
$\epsilon/N_c$ are neglected. At order $\epsilon$ the effective operators are obviously octets.
Explicit construction gives four basis operators, two  1-body and two 2-body operators.
Listed in Table \ref{tab1} are improved operators obtained by combining the octet pieces with singlet operators
in such a way that the resulting operators, $\bar{B}_{1,\cdots,4}$, have vanishing matrix elements between
non-strange baryons. Note that the operator $\bar{B}_2$ consists of two pieces of order $N_c$, namely $T^c_8$ and the 
operator $O_1$. However, the order $N_c$ pieces cancel and $\bar{B}_2$ is actually of order $N_c^0$.
It should be mentioned that, for reasons explained below,
several singlet and octet operators differ from those in \cite{CCGL} and \cite{SGS02}
by a scaling factor.

The calculation  of the matrix elements of these operators in the basis described in
Section II is a rather lengthy task. For this purpose  the Wigner-Eckart theorem has been repeatedly used
 to express any given matrix element in terms of combinations of $SU(2)$ and $SU(3)$
Clebsch-Gordan coefficients and the reduced matrix elements of the elementary operators
acting on the core and excited quark,  $\{S^c_i, T^c_a, G^c_{ia}\}$ and  $\{s_i,t_a,g_{ia}\}$,
respectively. As usual such reduced matrix elements were expressed in terms of
Clebsch-Gordan coefficients and some particular matrix elements which are simple
to evaluate explicitly. Fortunately, for the cases of  interest there exist
analytic formulas for all the $SU(2)$ coefficients\cite{Edm57} and $SU(3)$ isoscalar factors
\cite{Hec62,Ver68} involved in the calculations. Consequently, it has been possible
to derive explicit expressions, in terms of $N_c$, for all the matrix elements of the
singlet and octet operators included in the present analysis. Such expressions are
given in Tables \ref{tab2}-\ref{tab5}. It should be mentioned that all these formulas have been
independently checked by means of a numerical method based on a standard $SU(3)$ Wigner-coefficients package\cite{AD73}.
The matrix elements of the singlet operators in the sector of non-singlet states can also be checked
by comparing  with those obtained in the $SU(4)$ analysis in \cite{CCGL}.
Indeed, with the correspondences $O_i \to \tilde{O}_i$ $i=1,2,5-9$, and
$O_3\to \tilde{O}_3+\frac{1}{8} \tilde{O}_8$,
$O_4\to \tilde{O}_4+\frac{ N_c}{3(N_c+1)} \tilde{O}_5$, where the
$\tilde{O}_i$ are the operators in  $SU(4)$  differing with those in \cite{CCGL} by some of the
 rescaling factors that were mentioned  earlier,  it is  straightforward to cross-check the results.

It is also possible to check that certain combinations of operators are demoted to become of higher order in
$1/N_c$ in the sector of non-singlet baryons. As it was observed in \cite{CCGL} the combination of zeroth order
operators $O_2+O_4$ is of order $1/N_c$ and the combination of zeroth and first order operators
$O_2+O_4+\frac{2}{3} O_5+\frac{8}{3} O_9$ is of order $1/N_c^2$ in that sector.

\section{Fits and discussions}

In terms of the basis of operators introduced in the previous section
the {\bf 70}-plet  mass operator up to
order $1/N_c$ and order  $\epsilon N_c^0$  has the most general form:
\bea
M_{{\bf 70}} & =& \sum_{i=1}^{11} c_i O_i + \sum_{i=1}^{4} d_i \bar B_i~~~~,
\label{massform}
\eea
where $c_i$ and $d_i$ are the unknown real coefficients to be determined by fitting to the known masses and mixing
angles. These coefficients encode the  non-perturbative QCD dynamics that cannot be constrained by symmetries.
Calculating these coefficients would be equivalent to solving  QCD in this baryon sector. Fortunately, the
experimental data available in the case of the {\bf 70}-plet is enough to obtain them by performing  a
fit\cite{SGS02}. The inputs to the fit consist of seventeen masses of  negative parity baryons which have been assigned a status of three or
more stars by the Particle Data Group \cite{PDG},   and the two leading
order mixing angles $\theta_1= 0.61\pm 0.09$ and $\theta_3= 3.04\pm 0.15$ on which there is a rather good consensus
about their values as obtained from strong  decays of the non-strange members of the multiplet \cite{HLC75,IK1,CGO94}.
Note that   $\theta_3$ is consistent with zero mod $\pi$. Thus, the fifteen coefficients are fitted to these
nineteen observables. Of course, a larger number of inputs would have been desirable in order to increase the
confidence level of the results. It is however expected that the chief features of the results that are found here
are sufficiently  well established with these inputs.

Before proceeding to the numerical analysis of the masses, it is important to establish relations among observables
that must hold to the order of this analysis. Including $SU(3)$ breaking to all orders there are fifty observables,
namely thirty masses and twenty mixing angles.  Since there are fifteen operators in the basis being considered,
there must be thirty five relations. On the other hand, if $SU(3)$ breaking is considered only to order $\epsilon$,
there are thirty five observables where twenty one of them are masses and fourteen are mixing angles. This then
implies that up to order $\epsilon$ and order $1/N_c$ there are twenty relations. Among these relations there are those independent of the leading order mixing angles  resulting among traces of mass matrices for states with the same quantum numbers $I$ and $J$. There are thirteen such relations, which are independent of the coefficients $c_i$ and $b_i$, all of them involving mass splittings.  Five of them are Gell-Mann Okubo relations (one per  octet),
four equal spacing rules (two per decouplet), and four novel relations that involve mass splittings across $SU(3)$
multiplets. These relations are given by:
\bea
14 (s_{\Lambda_{3/2}}+ s_{\Lambda'_{3/2}}) + 63 s_{\Lambda_{5/2}}
+ 36 (s_{\Sigma_{1/2}}+ s_{\Sigma'_{1/2}}) &=& 68 (s_{\Lambda_{1/2}}+ s_{\Lambda'_{1/2}}) + 27 s_{\Sigma_{5/2}} \nonumber \\
14 (s_{\Sigma_{3/2}}+ s_{\Sigma'_{3/2}}) + 21 s_{\Lambda_{5/2}}-9 s_{\Sigma_{5/2}}
 &=& 18 (s_{\Lambda_{1/2}}+ s_{\Lambda'_{1/2}}) + 2 (s_{\Sigma_{1/2}}+
 s_{\Sigma'_{1/2}})
 \nonumber \\
14 s_{\Sigma''_{1/2}} + 49 s_{\Lambda_{5/2}} + 23 (s_{\Sigma_{1/2}}+
 s_{\Sigma'_{1/2}})
&=& 45 (s_{\Lambda_{1/2}}+ s_{\Lambda'_{1/2}}) + 19 s_{\Sigma_{5/2}}\nonumber \\
 14 s_{\Sigma''_{3/2}}  + 28 s_{\Lambda_{5/2}} + 11 (s_{\Sigma_{1/2}}+
 s_{\Sigma'_{1/2}})
&=&
 27 (s_{\Lambda_{1/2}}+ s_{\Lambda'_{1/2}}) + 10 s_{\Sigma_{5/2}},
\label{split}
\eea
where $s_{B_i}$ is the mass splitting between the state $B_i$ and the non-strange states in the $SU(3)$ multiplet to which it belongs. For the purpose of identifying the states within an $SU(3)$ multiplet the order $\epsilon$ mixing is disregarded. It is important to stress here that these relations are independent of the leading order as well as the order $\epsilon$ mixings.

As already mentioned, several of the singlet and octet operators defined in this article
differ from those in \cite{CCGL} and \cite{SGS02}
by a scaling factor. The reason for this is that here
the  operators have been defined in such a way that  their matrix elements are of the same
order in $1/N_c$ at which the operator contributes.  With this, the natural size of the coefficients
of singlet operators is about $500~{\rm MeV}$ as the analysis below shows, and that  of the coefficients  of $SU(3)$ breaking is about  $\epsilon\times 500~{\rm MeV}\sim 150-200 ~{\rm MeV}$.

The fit has been performed by treating the singlet pieces of the mass operator exactly and the $SU(3)$ breaking to
first order of perturbation theory in $\epsilon$. This approach  is justified in practice  by the fact that the
hyperfine interaction turns out to be the dominant spin-flavor breaking piece. The results of the fit are given in
Table~\ref{tab1}, where  the natural size of the coefficients associated with the singlet operators is seen to be
set by the coefficient of $O_1$. In fitting the data the experimental errors given by the Particle Data Group \cite{PDG}
are  taken whenever
they are larger than the expected magnitude of higher order corrections in the analysis. These  corrections are of
order $\epsilon^2$ or $\epsilon/N_c$, and their magnitude is taken to be about  $15~{\rm MeV}$. Although this is
not crucial for the outcome of the fit, the resulting   $\chi^2$ is more realistic. For instance, the singlet
$\Lambda$ masses are known experimentally within 5 MeV, and taking this error, which in magnitude would correspond
to a higher order of precision in the expansion in $\epsilon$ and $1/N_c$, would be unrealistic.

\begin{figure}[ht]
\begin{center}
\includegraphics[width=11cm,height=10cm]{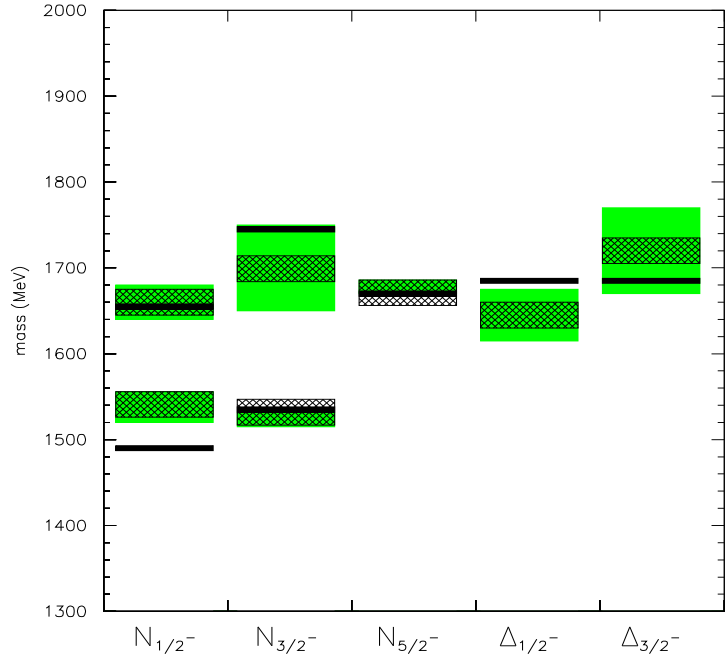}
\caption{ Non-strange baryon masses: the shaded boxes correspond to the experimental data{\protect
\cite{PDG}}, the black lines are the Isgur-Karl quark model predictions{\protect \cite{IK1}} and the
hatched boxes are the $1/N_c$ results{\protect \cite{SGS02}}.
\label{fig1} }
\end{center}
\end{figure}

\begin{figure}[ht]
\begin{center}
\includegraphics[width=10cm,height=10cm]{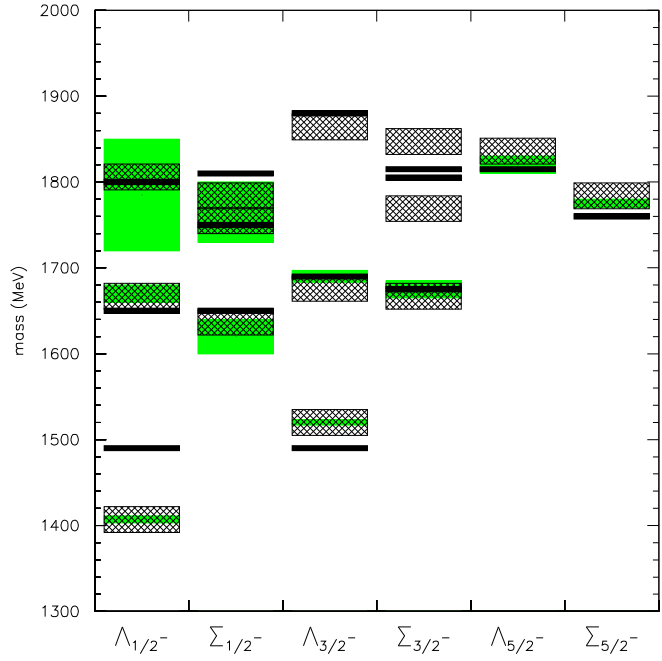}
\caption{ Masses of baryons with one unit of strangeness.
\label{fig2} }
\end{center}
\end{figure}

\begin{figure}[ht]
\begin{center}
\includegraphics[width=10cm,height=10cm]{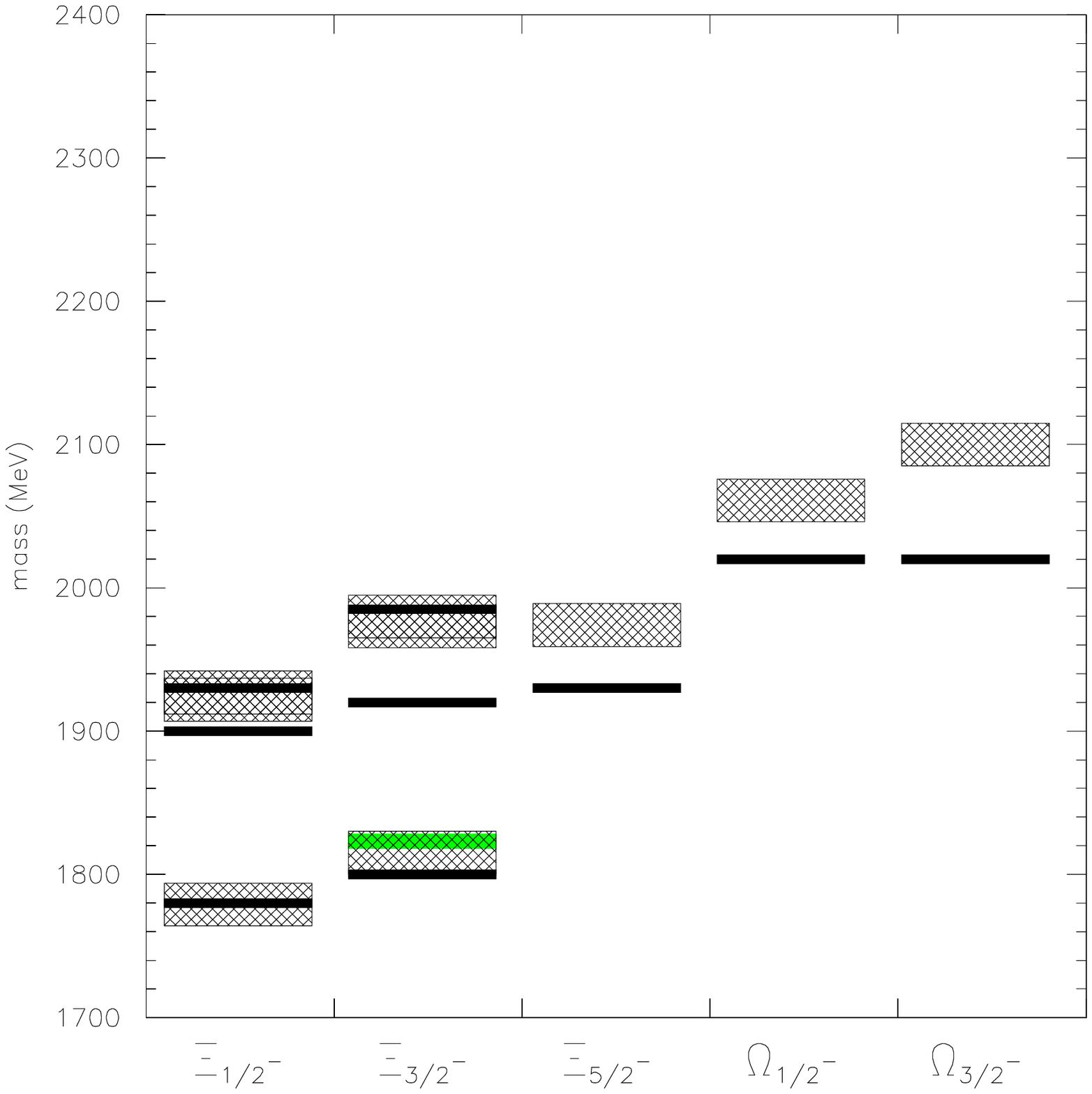}
\caption{Masses of baryons with two and three  units of strangeness.
\label{fig3} }
\end{center}
\end{figure}

Table~\ref{tab6} displays the empirical masses, and the masses and compositions of states resulting from the fit,
whose   $\chi^2$  per degree of freedom  turns out to be $1.29$. A clear display of the results is shown in Figs.
\ref{fig1}-\ref{fig3},
where the results are plotted along with the experimental values as well as the best fit provided by the
Isgur-Karl model\cite{IK1}.

In order to better understand the outcome of the fit, it is convenient to first emphasize the hierarchy that
emerges from the analysis. The contributions by the singlet operators to different mass matrix elements can be seen
by considering the contributions to the non-strange baryons and the singlet $\Lambda's$. Table \ref{tab7} explicitly
shows such
contributions. As it was already found in the analysis of the non-strange baryons \cite{CCGL}, the singlet
operators of order $N_c^0$ give contributions to the masses that are much smaller than the natural size expected at this order. Indeed, they turn out to  be of similar or even smaller magnitude than the natural size expected at  order $1/N_c$. Their importance however shows in the
leading order mixings driven by the matrix elements shown in the first two rows of Table \ref{tab7}. Among the operators
of order $1/N_c$, the dominant operator is the hyperfine operator $O_6$ that gives the chief spin-flavor breaking,
with all other operators giving contributions that are suppressed with respect to the  natural size. This
observed  hierarchy  that goes beyond the simple ordering in powers of $1/N_c$ reflects the
dynamics of QCD, and it is interesting to note that in its general aspects it agrees with the hierarchy that
results in constituent quark models.

A more detailed description of the role of the different singlet operators
is the following:

\begin{itemize}
\item The hyperfine operator $O_6$ gives the gross spin-flavor breaking features in the {\bf 70}-plet. This
operator does not affect  the singlet $\Lambda$ states which involve core states with $S^c=0$ only, while it
increases the masses of the rest of the states  according to the $S^c=1$ content of the cores. The typical mass
shifts produced by this operator are $160~{\rm MeV}$, which is the natural size expected from  $1/N_c$ counting
alone. In particular, this makes clear why the singlet $\Lambda$'s are the lightest states in the {\bf 70}-plet, as
it is well known from the early works in the constituent quark model based on QCD \cite{DGG75}. In the large $N_c$ limit the
hyperfine interaction in the core should be the same as in the ground state baryons. The $\Delta-N$ splitting gives
for the ground state baryons a strength of $100~{\rm MeV}$ for the hyperfine interaction defined by $\sum_{i\neq
j}^{N_c} s_i\cdot s_j$, while the corresponding strength implied by the result obtained for the coefficient $c_6$
is equal to $150~{\rm MeV}$. This disagreement is a manifestation of higher order corrections in $1/N_c$ and  is in line with the  expected magnitude for  $N_c=3$. It should be mentioned here that, as it occurs in the ground state baryons, the hyperfine operator has actually coherent matrix elements between states with spin of order $N_c$, for which the splittings
between states are of zeroth order. 

\item The spin-orbit operator $O_2$ has the peculiarity of being  one of the two operators that affect the
singlet $\Lambda$ masses (note that improved $SU(3)$ breaking operators also do, but
only through their singlet pieces proportional to the unity and spin-orbit operators, since the octet piece has
vanishing matrix elements for these states). The splitting between the singlet $\Lambda$'s is therefore a direct
measure of the spin-orbit interaction. The fact that the splitting is only $113~{\rm MeV}$ indicates the weakness
of the spin-orbit effect. This is perhaps the best indication that the formal problem originating in the fact that
spin-flavor symmetry is broken at order $N_c^0$ is rather harmless  in practice. Understanding the smallness of the
spin-orbit interaction from QCD is an important open dynamical problem.  In the constituent quark model picture a possible explanation
has been given in terms of the approximate cancellation between the spin-orbit pieces due to the Lorentz four-vector and scalar components of the effective potential \cite{Schnitzer,Ginocchio}. The sign of the spin-orbit splitting
between the  $\Lambda$ singlets obtained here
indicates that within that picture the Lorentz four-vector effective potential
should give
the larger contribution in  the case of the {\bf 70}-plet.

Note that the coefficient $c_2$ has a much smaller error than the combination  $c_2+c_4$ in
 the  non-strange sector analysis  \cite{CCGL} with which it ought to be compared. This is a consequence of the very important role of the singlet $\Lambda$ states that help pin down 
the contribution  of $O_2$ much more precisely.

One interesting observation can be made concerning the  splittings between spin-orbit partners in the {\bf 56}-plet. The positive parity
 {\bf 56}-plet with $\ell=2$ containing the spin-orbit partner  $N^*$ states $N_{3/2}(1720)$ and $N_{5/2}(1680)$, and the $\Delta^*$ states $\Delta_{1/2}(1910)$, $\Delta_{3/2}(1920)$, $\Delta_{5/2}(1905)$ and $\Delta_{7/2}(1950)$ \cite{PDG} shows very small  splittings. Indeed,  these splittings are suppressed by a factor one third or smaller with respect to  the spin-orbit splitting between the singlet $\Lambda$ states in the {\bf 70}-plet. Since the coupling of orbital angular momentum  within the {\bf 56}-plet is  of order $1/N_c$, it is possible that this  suppression is just a manifestation of the $1/N_c$ expansion.

\item The remaining two operators of order $N_c^0$ involve flavor exchange and give also contributions that are rather small, but important in the understanding of two issues.

i) The first issue is the so called spin-orbit puzzle in the quark model that can be summarized by the
incompatibility between such splittings in the sector of non-strange baryons and  in the singlet $\Lambda$'s. The
operator $O_4$ gives  contributions that compensate those of  $O_2$  to the splitting between the
spin-orbit partner states $\Delta_{1/2}$ and $\Delta_{3/2}$, where the manifestation of the spin-orbit puzzle was
most dramatic. The operators $O_3$ and $O_5$ do also give some relevant contributions to such splittings for other
states, but they are smaller than those of $O_4$. A  conclusion that can be drawn here is that the spin-orbit
puzzle in quark models is resolved by the  flavor-exchange effective interactions not included in that model and that appear naturally in the $1/N_c$ analysis.

ii) The second issue is the leading order mixings that are due to the off diagonal matrix elements in the two octet
mass matrices.  In
particular, the  mixing angle $\theta_1$, which is the only significant one of the two leading order angles,  is
almost entirely determined by the operator $O_3$, as the partial contributions displayed in Table \ref{tab7} show.
The angle $\theta_3$ receives contributions from several operators ($O_{2,\cdots,5,9,11}$) that are of similar
magnitude and tend to cancel. The first two rows in Table VII clearly show that the contributions to mixings by the order $1/N_c$ operators are small and even more tend to cancel each other to a substantial extent.

\item The hyperfine operator $O_7$ gives very small contributions. This operator involves the spin-spin interactions between the excited quark and the core, which  according to the constituent quark picture  is suppressed by the centrifugal barrier. As shown in Table VII, the current analysis shows that this operator gives splittings much smaller than those by $O_6$ and of the order of 25 MeV, in qualitative agreement with the quark picture.

\item The operator $O_5$ and the  three-body operators give contributions whose magnitude is in the few tens of
MeV, i.e. much smaller than the natural size of $1/N_c$ contributions, and have no clearly definite effect with
which they could be associated. They do however contribute to the ultimate quality of the fit. Finally, the
operator $O_8$ is clearly irrelevant.

\item Table \ref{tab8} shows the results of partial fits to the masses of the
non-strange  and the singlet $\Lambda$  baryons listed in the first column of Table \ref{tab7}.
The contribution of the strange quark to the $\Lambda$ masses has been taken to
be $135~{\rm MeV}$ as it results from the average empirical splitting per unit of strangeness.
The progress in the quality of the fit can be observed  as operators are sequentially included.
Comparing the fitted coefficients with the values obtained from using the
full basis $O_{1,...,11}$, {\it i.e.} those listed in the last line of Table \ref{tab8} (note that these values can be exactly obtained since there are eleven observables)
it can be noticed that the values obtained in successive partial fits always remain close to them.

\end{itemize}

Concerning $SU(3)$ breaking, only three out of the four  $SU(3)$ breaking operators are significant giving
natural size contributions.  $\bar{B}_3$ is weak and can to some extent be disregarded. The dominance of the $O_6$
operator  may indicate that  associated octet operators such as $\frac{1}{N_c} S^c_i G^c_{i8}$, which is of order
$\epsilon/N_c$, would be important, even when it appears at higher order than the ones considered in this paper.
Although this may be so,  the available data for splittings does not allow to pin down the relevance of such an
operator. The fact is that with the four improved leading order operators already included    the fit is  very
good, and the inclusion of  such operator does not lead to a  significant improvement. More data on $SU(3)$
splittings would be required to clarify this issue. The inclusion of such an operator would spoil the splitting
relations of Eq. (\ref{split}). The main observations on $SU(3)$ breaking are the following:

\begin{itemize}

\item Only one relation can be tested with available data, namely the Gell-Mann Okubo relation for the $J=3/2$
octet that is predominantly $S=1/2$. Determining the masses of the $\Xi'_{1/2}$ and  $\Xi_{5/2}$ would complete two
more octets and test the corresponding relations. A further octet can be completed  that  has one two-star state,
the $\Sigma_{1/2}(1620)$ state, by finding the corresponding $\Xi$ state predicted by the analysis to have a mass
of $1779~{\rm MeV}$.

\item The test of equal spacing relations is not possible. Only the $\Delta$ states in the two decouplets are
known. It is clearly very important  at some point in the future to have further decouplet states experimentally pinned
down for that purpose. The results obtained here indicate that the  splittings in both decouplets are similar and
in the range $125-135~{\rm MeV}$, which is the typical splitting produced by one unit of strangeness.

\item The Gell-Mann Okubo and equal spacing relations are violated at order $\epsilon^2$, while the four relations
Eq. (\ref{split}) are violated at order $\epsilon/N_c$ and $\epsilon^2$. While in the ground state baryons the  Gell-Mann
Okubo and equal spacing relations are  violated at order $\epsilon^2/N_c$ \cite{JL95}, the order $\epsilon^2$
violation of the relations in excited baryons results primarily from the $SU(3)$ breaking mixings between $\Sigma$s
and between $\Lambda$s belonging to different multiplets.

\item The four new splitting relation of  Eq.(\ref{split}) cannot be tested  at this point because the masses of
$\Lambda'_{3/2}$, $\Sigma'_{3/2}$, $\Sigma''_{1/2}$, and $\Sigma''_{1/2}$ that enter respectively in the four
relations need to be known. After replacing the known experimental values, each of the splitting relations
gives a prediction, namely: $ s_{\Lambda'_{3/2}}= 149 \ {\rm MeV} $, $ s_{\Sigma'_{3/2}} =  55 \ {\rm MeV} $, $
s_{\Sigma''_{1/2}}= 154 \ {\rm MeV} $, and $ s_{\Sigma''_{3/2}}= 134 \ {\rm MeV} $.  If the operator $\bar{B}_3$ is
ignored, five relations result that were given in \cite{SGS02}. In that case one relation could be tested, namely
the relation $9 (s_{\Sigma_{1/2}} +  s_{\Sigma'_{1/2}}) + 21 s_{\Lambda_{5/2}} = 17 (s_{\Lambda_{1/2}} +
s_{\Lambda'_{1/2}}) + 5 s_{\Sigma_{5/2}}$; by  including the two-star state $\Sigma_{1/2}(1620)$ as input that
relation is satisfied to a few percent.

\end{itemize}

\section{Conclusions}

The $1/N_c$ expansion for excited baryons has been implemented under the assumption that there is an approximate
spin-flavor symmetry in the large $N_c$ limit.
This assumption relies on the observation that zeroth order violations of this symmetry are
very small in practice. Consequently,  the only effects that have been left out in the analysis carried out for the $SU(6)$ ${\bf 70-}$plet
are related to spin-flavor  configuration mixing. Since the analysis shows that the zeroth order spin-flavor
breaking in the ${\bf 70-}$plet has a magnitude smaller than the natural size of first order contributions, the scheme is phenomenologically sound.

The analysis also shows that the $1/N_c$ expansion can be consistently applied because there are no corrections that are
unnaturally large. On the other hand, a hierarchy emerges  in the form of effective coefficients being unnaturally
small. In its gross features the picture that emerges is similar to the quark model one, but at a finer level  the
suppressed dynamics  manifests itself  in particular through flavor exchange effective interactions, largely absent
in most quark models,  which are important in describing two effects, namely, the zeroth order mixings and the
resolution  of the spin-orbit puzzle.

At the level of $SU(3)$ singlet spin-flavor symmetry breaking the level of predictivity is quite limited, the
reason being that the number of observables is equal to the number of operators in the singlet basis up to order
$1/N_c$. There is however predictivity at the level of $SU(3)$ breaking to order $\epsilon\times N_c^0$: besides
Gell-Mann Okubo and equal spacing relations there are four new relations across different $SU(3)$ multiplets.
Unfortunately, with the available data only one Gell-Mann Okubo relation can be tested. This should be a
motivation to experimentally establish a few more key states in the  ${\bf 70-}$plet.

The present analysis  provides  a useful framework to sort out and understand results from lattice QCD simulations of excited baryons. The $1/N_c$ expansion allows to separate the contributions that follow from the dynamical $SU(2F)$ symmetry
and its breaking, from the non-perturbative reduced matrix elements of the QCD operators. In particular, the
$\Lambda (1405)$ appears naturally as the lightest state and a spin-orbit partner of the $\Lambda (1520)$.
The spin-spin and spin-orbit interactions that give the gross structure of the ${\bf 70}$-plet  are especially
interesting and lattice simulations together with the $1/N_c$ analysis could help to further understand their nature.

\section*{Acknowledgments}

The authors would like to thank  Wally Melnitchouk, David Richards and Yuri Simonov for useful remarks, and Winston Roberts for enlightening discussions and comments on the manuscript. This work was partially supported by the National Science Foundation (USA) through grant \#~PHY-9733343
(JLG,CLS), by the ANPCYT (Argentina) through grant \#~03-08580 (NNS) and by  sabbatical leave support  from SURA (JLG).
Two  of us (JLG and CLS) thank  the Institut f\"ur Theoretische Physik of the University of Bern for the
kind hospitality and partial support while part of this work was completed. This work was supported by DOE
contract DE-AC05-84ER40150 under which SURA operates the Thomas Jefferson National Accelerator Facility and 
also partially supported by DOE grant DE-FG02-96ER40945.
Support from Fundaci\'on Antorchas (Argentina) is also acknowledged.

\pagebreak


\begin{table}[ht]
\caption{ List of operators and the coefficients resulting from the best fit to the
known  {\bf 70}-plet masses and mixings {\protect \cite{SGS02}}.}
\label{tab1}
\begin{center}
\footnotesize
\begin{tabular}{llrrr}
\hline
\hline
Operator & \multicolumn{4}{c}{Fitted coef. [MeV]}\\
\hline
\hline
$O_1 = N_c \ 1 $ & $c_1 =$  & 449 & $\pm$ & 2 $\ $  \\
\hline
$O_2 = l_i \ s_i$ & $c_2 =$ & 52 & $\pm$ & 15   $\ $ \\
$O_3 = \frac{3}{N_c} \ l^{(2)}_{ij} \ g_{ia} \ G^c_{ja} $ & $c_3 =$  & 116 & $\pm$ & 44  $\ $ \\
$O_4 = \frac{4}{N_c+1} \ l_i \ t_a \ G^c_{ia}$ & $c_4 =$  & 110 & $\pm$ &  16 $\ $\\
\hline
$O_5 = \frac{1}{N_c} \ l_i \ S^c_i$ & $c_5 =$  & 74 & $\pm$ & 30 $\ $\\
$O_6 = \frac{1}{N_c} \ S^c_i \ S^c_i$ & $c_6 =$  & 480 &  $\pm$ & 15 $\ $\\
$O_7 = \frac{1}{N_c} \ s_i \ S^c_i$ & $c_7 =$ & -159 &  $\pm$ & 50 $\ $ \\
$O_8 = \frac{2}{N_c} \ l^{(2)}_{ij} s_i \ S^c_j$ & $c_8  =$  & 3  & $\pm$ & 55   $\ $\\
$O_9 = \frac{3}{N_c^2} \ l_i \ g_{ja} \{ S^c_j ,  G^c_{ia} \} $ & $c_9 =$ &  71 &  $\pm$ &  51  $\ $\\
$O_{10} = \frac{2}{N_c^2} t_a \{ S^c_i ,  G^c_{ia} \}$ & $c_{10} =$  & -84 &  $\pm$ &  28  $\ $\\
$O_{11} = \frac{3}{N_c^2} \ l_i \ g_{ia} \{ S^c_j ,  G^c_{ja} \}$ & $c_{11} =$ & -44 &  $\pm$ &  43  $\ $\\
\hline
\hline
$\bar B_1 = t_8 - \frac{1}{2 \sqrt{3} N_c} O_1$ & $d_1 =$  & -81 & $\pm$ & 36 $\ $\\
$\bar B_2 = T_8^c - \frac{N_c-1}{2 \sqrt{3} N_c } O_1 $  & $ d_2 = $  & -194 & $\pm$ & 17  $\ $\\
$\bar B_3 = \frac{10}{N_c} \  d_{8ab}  \ g_{ia} \ G^c_{ib}  + \frac{5(N_c^2 -9)}{8 \sqrt{3} N_c^2 (N_c-1)} O_1 +$ &  & & $\ $\\
\hspace*{1cm} $+ \frac{5}{2 \sqrt{3} (N_c-1)} O_6 + \frac{5}{6 \sqrt{3}} O_7 $  & $ d_3 = $  & -15 & $\pm$ & 30  $\ $\\
$\bar B_4 =3\ l_i \ g_{i8} - \frac{\sqrt{3}}{2} O_2 $ & $ d_4 = $  & -27 & $\pm$ & 19  $\ $\\
\hline \hline
\end{tabular}
\end{center}
\end{table}


\renewcommand{\arraystretch}{1.0}
{\squeezetable
\begin{table}[ht]
\caption{Matrix elements of singlet operators $O_1$ to $O_6$ in the {\bf 70}-plet. Note that although it is not explicitly
indicated, within each subspace the matrix elements which are not diagonal with respect to
isospin $I$ and strangeness $S$ vanish.}
\label{tab2}
\begin{tabular}{ccccccc}\\
\hline\hline
                     & $O_1$ &  $O_2$ & $O_3$ & $O_4$ & $O_5$ & $O_6$  \\[1.mm]
                      &
                $N_c \ 1$                   &
               $l_i\ s_i$                  &
$\frac{3}{N_c}\ l^{(2)}_{ij} \ g_{ia} \ G^c_{ja}$ &
$\frac{4}{N_c+1}\ l_i \ t_a \ G^c_{ia} $         &
$\frac{1}{N_c}\ l_i \ S^c_i$                   &
$\frac{1}{N_c}\ S^c_i \ S^c_i$                 \\
                     \hline
$^28_{\frac{1}{2}}$            &
          $N_c$   &
          $ \frac{3-2 N_c}{3 N_c} $ &
          $0$ &
          $\frac{2}{9}  \frac{(N_c + 3)(3 N_c-2)}{N_c(N_c+1)}$ &
          $-\frac{N_c+3}{3N_c^2} $ &
          $\frac{N_c+3}{2N_c^2} $ \\ [1.mm]
$^48_{\frac{1}{2}}$           &
          $N_c$   &
          $-\frac{5}{6} $ &
          $ -\frac{5(3 N_c + 1 )}{48 N_c}$ &
          $\frac{5(3 N_c + 1)}{18(N_c+1)}  $  &
          $-\frac{5}{3 N_c}$ &
          $\frac{2}{N_c}$ \\ [1.mm]
$^28_{\frac{1}{2}} - ^48_{\frac{1}{2}}$ &
          $0$   &
          $-\frac{1}{3 \sqrt{2}} \sqrt{ 1 + \frac{3}{N_c}}$ &
          $-\frac{5(3 N_c-2)}{24 \sqrt{2} N_c} \sqrt{1 + \frac{3}{N_c} }$ &
          $-\frac{1}{9\sqrt{2}} \frac{ 5-3 N_c }{N_c+1}  \sqrt{1 + \frac{3}{N_c} }$ &
          $\frac{1}{3\sqrt{2}N_c}  \sqrt{1 + \frac{3}{N_c} } $ &
          $0$\\ [1.mm]
$^21_{\frac{1}{2}}$            &
          $N_c$   &
          $ -1 $ &
          $0$ &
          $0$ &
          $0$ &
          $0$  \\ [1.mm] 
$^210_{\frac{1}{2}}$           &
          $N_c$   &
          $\frac{1}{3}$ &
          $0$  &
          $-\frac{(3 N_c + 7)}{9(N_c + 1)} $ &
          $- \frac{4}{3 N_c}$ &
          $\frac{2}{N_c}$\\ [1.mm] \hline
$^28_{\frac{3}{2}}$            &
          $N_c$   &
          $\frac{2 N_c - 3}{6 N_c}$ &
          $0$ &
          $ -\frac{1}{9} \frac{(N_c+3)(3 N_c-2)}{N_c(N_c+1)} $ &
          $\frac{N_c+3}{6N_c^2} $ &
          $\frac{N_c+3}{2N_c^2} $ \\ [1.mm]
$^48_{\frac{3}{2}}$           &
          $N_c$   &
          $-\frac{1}{3} $ &
          $\frac{ 3 N_c + 1}{12 N_c}$ &
          $\frac{3 N_c + 1}{9( N_c + 1)}$ &
          $- \frac{2}{3 N_c}$ &
          $\frac{2}{N_c}$\\ [1.mm]
$^28_{\frac{3}{2}} - ^48_{\frac{3}{2}}$ &
          $0$     &
          $-\frac{\sqrt{5}}{6}  \sqrt{1 + \frac{3}{N_c} }$ &
          $-\frac{\sqrt{5}(2-3 N_c )}{48 N_c} \sqrt{1 + \frac{3}{N_c} } $ &
          $-\frac{\sqrt{5}}{18} \frac{ 5-3 N_c }{N_c+1}  \sqrt{1 + \frac{3}{N_c} } $ &
          $\frac{\sqrt{5}}{6 N_c}  \sqrt{1 + \frac{3}{N_c} }$&
          $0$ \\ [1.mm]
$^21_{\frac{3}{2}}$            &
          $N_c$   &
          $\frac{1}{2}$ &
          $0$ &
          $0$ &
          $0$ &
          $0$ \\ [1.mm]
$^210_{\frac{3}{2}}$           &
          $N_c$   &
          $-\frac{1}{6}$ &
          $0$ &
          $\frac{3 N_c + 7}{18(N_c + 1)} $ &
          $\frac{2}{3 N_c}$ &
          $\frac{2}{N_c}$ \\[1.mm] \hline
$^48_{\frac{5}{2}}$            &
          $N_c$   &
          $\frac{ 1}{2}$ &
          $-\frac{3 N_c + 1}{48 N_c} $ &
          $-\frac{3 N_c + 1}{6(N_c + 1)}$ &
          $\frac{1}{N_c}$ &
          $\frac{2}{N_c}$\\ [2mm]  \hline\hline
\end{tabular}

\end{table}}

\pagebreak


{\squeezetable
\begin{table}[ht]
\caption{Matrix elements of singlet operators $O_7$ to $O_{11}$ in the {\bf 70}-plet. Note that although it is not explicitly
indicated, within each subspace the matrix elements which are not diagonal with respect to
isospin $I$ and strangeness $S$ vanish.}
\label{tab3}
\begin{tabular}{cccccc}\\
\hline\hline
                     &  $O_7$ & $O_8$ & $O_9$ & $O_{10}$ & $O_{11}$  \\[1.mm]
                     &
$\frac{1}{N_c}\ s_i S^c_i$                      &
\hspace{3.mm}$\frac{2}{N_c}\ l^{(2)}_{ij} \ s_{i} \ S^c_{j}$ \hspace{3.mm}&
\hspace{3.mm} $\frac{3}{N_c^2} \ l_i \ g_{ja} \ \{S^c_j, G^c_{ia} \}$ \hspace{3.mm}&
\hspace{3.mm} $\frac{2}{N_c^2} \ t_{a} \ \{S^c_i, G^c_{ia} \}$ \hspace{3.mm} &
\hspace{3.mm} $\frac{3}{N_c^2} \ l_i \ g_{ia} \ \{S^c_j, G^c_{ja} \}$ \hspace{3.mm}\\
                     \hline
$^28_{\frac{1}{2}}$            &
          $ -\frac{(N_c + 3)}{4N_c^2}  $ &
          $0$ &
          $\frac{(N_c+3)(7-15 N_c )}{24 N_c^3}$ &
          $-\frac{(N_c+3)(3 N_c +1)}{12 N_c^3}$ &
          $-\frac{(N_c+3)(3 N_c +1)}{24 N_c^3}$ \\ [1.mm]
$^48_{\frac{1}{2}}$           &
          $\frac{1}{2 N_c}$ &
          $\frac{5}{3 N_c}$ &
          $\frac{5(3 N_c + 1)}{24 N_c^2} $ &
          $-\frac{(3 N_c + 1)}{3 N_c^2} $ &
          $\frac{5(3 N_c + 1)}{12 N_c^2} $ \\ [1.mm]
$^28_{\frac{1}{2}} - ^48_{\frac{1}{2}}$ &
          $0$ &
          $\frac{5}{6 \sqrt{2} N_c} \sqrt{1 + \frac{3}{N_c} }$ &
          $\frac{3 N_c-2}{12\sqrt{2} N_c^2} \sqrt{1 + \frac{3}{N_c} }$ &
          $0$ &
          $\frac{3 N_c + 1}{6 \sqrt{2} N_c^2} \sqrt{1 + \frac{3}{N_c}}$ \\ [1.mm]
$^21_{\frac{1}{2}}$            &
          $0$ &
          $0$ &
          $0$ &
          $0$ &
          $0$ \\ [1.mm]
$^210_{\frac{1}{2}}$           &
          $- \frac{1}{N_c}$ &
          $0$  &
          $\frac{(3 N_c + 7)}{6 N_c^2}$ &
          $\frac{(3 N_c + 7)}{6 N_c^2}$ &
          $\frac{(3 N_c + 7)}{12 N_c^2}$ \\ [1.mm] \hline
$^28_{\frac{3}{2}}$            &
          $ -\frac{(N_c + 3)}{4N_c^2} $ &
          $0$ &
          $\frac{(N_c+3)(15 N_c- 7)}{48  N_c^3}$ &
          $-\frac{(N_c+3)(3 N_c +1)}{12  N_c^3}$  &
          $\frac{(N_c+3)(3 N_c +1)}{48 N_c^3}$\\ [1.mm]
$^48_{\frac{3}{2}}$           &
          $\frac{1}{2 N_c} $ &
          $- \frac{4}{3 N_c}$ &
          $\frac{(3 N_c + 1)}{12 N_c^2}$ &
          $-\frac{(3 N_c + 1)}{3 N_c^2} $ &
          $\frac{(3 N_c + 1)}{6 N_c^2} $ \\ [1.mm]
$^28_{\frac{3}{2}} - ^48_{\frac{3}{2}}$ &
          $0$ &
          $-\frac{\sqrt{5}}{12 N_c}          \sqrt{1 + \frac{3}{N_c}}  $ &
          $\frac{\sqrt{5} (3 N_c-2 )}{24 N_c^2} \sqrt{1 + \frac{3}{N_c}}  $ &
          $0$ &
          $\frac{\sqrt{5}(3 N_c + 1)}{12  N_c^2} \sqrt{1 + \frac{3}{N_c}}$ \\ [1.mm]
$^21_{\frac{3}{2}}$            &
          $0$ &
          $0$ &
          $0$ &
          $0$ &
          $0$ \\ [1.mm]
$^210_{\frac{3}{2}}$           &
          $-\frac{1}{N_c}$ &
          $0$ &
          $-\frac{(3 N_c + 7)}{12 N_c^2}$ &
          $ \frac{(3 N_c + 7)}{6 N_c^2}$ &
          $-\frac{(3 N_c + 7)}{24 N_c^2}$ \\[1.mm] \hline
$^48_{\frac{5}{2}}$            &
          $\frac{1}{2 N_c}$ &
          $\frac{1}{3 N_c}$ &
          $-\frac{(3 N_c + 1)}{8 N_c^2}$ &
          $-\frac{(3 N_c + 1)}{3 N_c^2} $ &
          $-\frac{(3 N_c + 1)}{4 N_c^2}$ \\ [2mm] \hline\hline
\end{tabular}
\end{table}}

\pagebreak


\begin{turnpage}
{\squeezetable
\begin{table}[ht]
\caption{Matrix elements of isospin-singlet octet operators in the {\bf 70}-plet.}
\label{tab4}
\begin{tabular}{cccc}
\\
\hline\hline
                     &  $B_1$ & $B_2$ & $B_3$  \\[1.mm]
                     &
  $t_8$              &
  $T^c_8$            &
 $\frac{10}{N_c}\ d_{8ab} \ g_{ia} \ G^c_{ib}$ \\
\hline\\[-3.mm]
$^28_{\frac{1}{2}}$,
$^28_{\frac{3}{2}}$
& $  \frac{ N_c^3 + (7 S+ 8 I^2)N_c^2 - 3(4 S + 8 I^2 -1) N_c + 9 S}{2\sqrt{3} N_c(N_c-1)(N_c+3)}$& $  \frac{ N_c^4 + (3 S+ 1)N_c^3 - (S + 8 I^2 + 3) N_c^2 + 3 (S + 8I^2-1)N_c - 9 S}{2\sqrt{3} N_c(N_c-1)(N_c+3)}$
& $ \frac{3 N_c^3 + (13 S + 8 I^2 -3) N_c^2 - (31 S + 44 I^2 +12)N_c + 6 ( S + 14 I^2)}{-\frac{24}{5}\sqrt{3} N_c^2(N_c-1)}$
\\[1mm]
$^48_{\frac{1}{2}}$ ,
$^48_{\frac{3}{2}}$ ,&&&\\
$^48_{\frac{5}{2}}$  &
   $\frac{N_c+S-4 I^2}{2\sqrt{3}(N_c-1)}$ &
   $\frac{N_c^2 + ( 3 S -2) N_c + 4 (I^2-S) }{2\sqrt{3}(N_c-1)}$ &
   $\frac{3 N_c^2 + (7 S - 4 I^2 + 3) N_c - ( S + 20 I^2)}{-\frac{24}{5}\sqrt{3} N_c(N_c-1)}$ \\[1mm]
$^28_{\frac{1}{2}} - ^48_{\frac{1}{2}}$ ,&&&\\
$^28_{\frac{3}{2}} - ^48_{\frac{3}{2}}$
       &
   $0$ &
   $0$ &
   $0$ \\[1mm]
$^21_{\frac{1}{2}}$ ,
$^21_{\frac{3}{2}}$
                  &
   $ \frac{(3-N_c)}{\sqrt{3} (N_c+3)}$ &
   $ \frac{(N_c+5)(N_c-3)}{2 \sqrt{3} (N_c+3)}$ &
   $0$ \\[1mm]
$^28_{\frac{1}{2}}-^21_{\frac{1}{2}}$,&&&\\
$^28_{\frac{3}{2}}-^21_{\frac{3}{2}}$
                                  &
   $-\frac{3 (N_c-1)}{2 \sqrt{N_c} (N_c+3)}$  &
   $- \frac{3(N_c-1)}{2 \sqrt{N_c} (N_c+3)}$  &
   $\frac{5(3 N_c + 1)}{16 N_c \sqrt{N_c} }$   \\
$^48_{\frac{1}{2}}-^21_{\frac{1}{2}}$ ,&&&\\
$^48_{\frac{3}{2}}-^21_{\frac{3}{2}}$
                                   &
   $0$ &
   $0$ &
   $0$ \\[1mm]
$10_{\frac{1}{2}}$ ,
$10_{\frac{3}{2}}$
                   &
   $ \frac{N_c + 8 S + 5}{2 \sqrt{3} (N_c+5)}$ &
   $ \frac{N_c^2 + (3 S + 4) N_c + 7 S - 5}{2 \sqrt{3} (N_c+5)}$ &
   $ - \frac{3 N_c^2 + 14 (S + 1) N_c + 22 S - 5}{\frac{24}{5} \sqrt{3} N_c (N_c+5)}$ \\
$^28_{\frac{1}{2}}-^210_{\frac{1}{2}}$, &&&\\
$^28_{\frac{3}{2}}-^210_{\frac{3}{2}}$
                                   &
   $-\sqrt{\frac{2}{3}} \sqrt{\frac{N_c+3}{N_c(N_c-1)(N_c+5)}}$ &
   $\sqrt{\frac{2}{3}} \sqrt{\frac{N_c+3}{N_c(N_c-1)(N_c+5)}}$  &
   $\frac{5(N_c+2)}{6 \sqrt{6} N_c} \sqrt{\frac{N_c+3}{N_c(N_c-1)(N_c+5)}}$ \\
$^48_{\frac{1}{2}}-^210_{\frac{1}{2}}$, &&&\\
$^48_{\frac{3}{2}}-^210_{\frac{3}{2}}$
                                    &
   $0$ &
   $0$ &
   $0$ \\ [2mm]
\hline\hline
\end{tabular}
\end{table}}
\end{turnpage}

\pagebreak


\begin{table}[ht]
\caption{Matrix elements of the isospin-singlet  octet operator $B_4$ in the {\bf 70}-plet.}
\label{tab5}
\begin{tabular}{cc}\\
\hline\hline
                     & $B_4$ \\[1mm]
                     &
 $3 l_i \ g_{i8}$ \\
    \hline
$^28_{\frac{1}{2}}$ &
$ - \frac{ N_c^3 + (10 S + 14 I^2 -3) N_c^2 - 3 (7 S + 8I^2)N_c + 9 ( S + 2 I^2)}{\sqrt{3} N_c(N_c-1)(N_c+3)}$ \\
$^48_{\frac{1}{2}}$ &
   $- \frac{5(N_c + S - 4 I^2)}{4\sqrt{3}(N_c-1)}$ \\
$^28_{\frac{1}{2}} - ^48_{\frac{1}{2}}$ &
   $-\frac{N_c+S-4I^2}{2\sqrt{6} (N_c-1)} \sqrt{ 1 + \frac{3}{N_c}}$  \\
$^21_{\frac{1}{2}}$ &
   $\frac{\sqrt{3}(N_c-3)}{ (N_c+3)}$  \\
$^28_{\frac{1}{2}}-^21_{\frac{1}{2}}$ &
   $\frac{9 (N_c-1)}{2 (N_c+3)\sqrt{N_c} }$ \\
$^48_{\frac{1}{2}}-^21_{\frac{1}{2}}$ &
   $0$  \\
$^210_{\frac{1}{2}}$ &
   $   \frac{N_c + 8 S + 5}{2 \sqrt{3} (N_c+5)}$  \\
$^28_{\frac{1}{2}}-^210_{\frac{1}{2}}$ &
   $ -\sqrt{\frac{2}{3}} \sqrt{\frac{N_c+3}{N_c(N_c-1)(N_c+5)}}$ \\
$^48_{\frac{1}{2}}-^210_{\frac{1}{2}}$ &
   $\frac{4}{\sqrt{3}} \frac{1}{\sqrt{(N_c-1)(N_c+5)}}$  \\ \hline
$^28_{\frac{3}{2}}$ & \
 $ \frac{N_c^3 + (10 S + 14 I^2 -3) N_c^2 - 3(7 S + 8 I^2) N_c + 9( S + 2 I^2)}{ 2 \sqrt{3} N_c (N_c-1) (N_c+3) }$  \\
$^48_{\frac{3}{2}}$ &
   $ - \frac{ N_c + S - 4 I^2 }{ 2 \sqrt{3} (N_c-1) } $ \\
$^28_{\frac{3}{2}} - ^48_{\frac{3}{2}}$ &
   $-\sqrt{\frac{5}{3}} \frac{N_c+S-4I^2}{4(N_c-1)} \sqrt{ 1 + \frac{3}{N_c}} $\\
$^21_{\frac{3}{2}}$ &
   $-  \frac{\sqrt{3}(N_c-3)}{2(N_c+3)}$ \\
$^28_{\frac{3}{2}}-^21_{\frac{3}{2}}$ &
    $ -\frac{9(N_c-1)}{4(N_c+3)\sqrt{N_c}}$ \\
$^48_{\frac{3}{2}}-^21_{\frac{3}{2}}$ &
  $0$  \\
$^210_{\frac{3}{2}}$ &
  $- \frac{N_c+8 S + 5}{4 \sqrt{3} (N_c+5)}$   \\
$^28_{\frac{3}{2}}-^210_{\frac{3}{2}}$ &
   $   \sqrt{\frac{N_c+3}{6N_c(N_c-1)(N_c+5)} }$ \\
$^48_{\frac{3}{2}}-^210_{\frac{3}{2}}$ &
  $\frac{2 \sqrt{10}}{\sqrt{3(N_c-1)(N_c+5)} }$  \\ \hline
$^48_{\frac{5}{2}}$ &
   $\frac{\sqrt{3}(N_c+S-4 I^2)}{4(N_c-1)}$
\\ [2mm]
\hline\hline
\end{tabular}

\vspace*{.5cm}

\end{table}


\renewcommand{\arraystretch}{.8}
\begin{table}[ht]
\caption{\small Masses and spin-flavor content as predicted by the  $1/N_c$ expansion{\protect \cite{SGS02}} respectively depicted in the third and last four columns . The empirical values of the  masses for all states with a status of three or more stars in \cite{PDG} are shown in the second column and the results from the  quark model
 calculation of Isgur and Karl {\protect \cite{IK1}} are shown in the fourth column. }
\label{tab6}
\begin{center}
\footnotesize
\begin{tabular}{ccccccccccc}\hline \hline
       & & \multicolumn{3}{c}{Masses [MeV]} & &  \multicolumn{4}{c}{Spin-flavor content} \\
State &\hspace*{.5cm} & Expt. & Large $ N_c$  &  QM &\hspace*{.5cm} & $^21$ & $^28$ & $^48$ & $^210$ \\
\hline
  $N_{1/2}$      & &$ 1538 \pm 18  $ & 1541  & 1490  &  &               &  0.82        &  0.57  &         \\
 $\Lambda_{1/2}$ & &$ 1670 \pm 10  $ & 1667  & 1650  &  & -0.21  &  0.90  &  0.37  &         \\
 $\Sigma_{1/2}$  & &$  (1620)      $ & 1637  & 1650  &  &               &  0.52  &  0.81  &  0.27   \\
 $\Xi_{1/2}$     & &$              $ & 1779  & 1780  &  &       &  0.85  &  0.44 &  0.29  \\
\hline
 $N_{3/2}  $     & &$ 1523 \pm 8   $ & 1532  & 1535  &  &       &-0.99   & 0.10   &         \\
 $\Lambda_{3/2}$ & &$ 1690 \pm 5   $ & 1676  & 1690  &  & 0.18  & -0.98  & 0.09   &         \\
 $\Sigma_{3/2}$  & &$ 1675 \pm 10  $ & 1667  & 1675  &  &       & -0.98  & -0.01   &-0.19 \\
 $\Xi_{3/2}$     & &$ 1823 \pm 5   $ & 1815  & 1800  &  &       & -0.98  &  0.03  & -0.19   \\
\hline
 $N'_{1/2}  $    & &$ 1660 \pm 20  $ & 1660  & 1655  &  &       & -0.57  &  0.82  &         \\
 $\Lambda'_{1/2}$& &$ 1785 \pm 65  $ & 1806  & 1800  &  & 0.10  & -0.38  & 0.92  &         \\
 $\Sigma'_{1/2}$ & &$ 1765 \pm 35  $ & 1755  & 1750  &  &       & -0.83  &  0.54  &  0.17   \\

 $\Xi'_{1/2}$    & &$              $ & 1927  & 1900  &  &       & -0.46  &  0.87  &  0.18   \\
\hline
 $N'_{3/2}  $    & &$ 1700 \pm 50  $ & 1699  & 1745  &  &       & -0.10  & -0.99  &         \\
 $\Lambda'_{3/2}$& &$              $ & 1864  & 1880  &  &0.01   & -0.09  & -0.99  &         \\
 $\Sigma'_{3/2}$ & &$              $ & 1769  & 1815  &  &       & 0.01   & (-0.57)  & (-0.82)   \\
 $\Xi'_{3/2}$    & &$              $ & 1980  & 1985  &  &       & -0.02  & (-0.57)  & (-0.82)   \\
\hline
 $N_{5/2}  $     & &$ 1678 \pm 8   $ & 1671  & 1670  &  &       &        &  1.00  &         \\
 $\Lambda_{5/2}$ & &$ 1820 \pm 10  $ & 1836  & 1815  &  &       &        &  1.00  &         \\
 $\Sigma_{5/2}$  & &$ 1775 \pm 5   $ & 1784  & 1760  &  &       &        &  1.00  &         \\
 $\Xi_{5/2}$     & &$              $ & 1974  & 1930  &  &       &        &  1.00  &         \\
\hline
 $\Delta_{1/2}$  & &$ 1645 \pm 30  $ & 1645  & 1685  &  &       &        &        &  1.00   \\
 $\Sigma''_{1/2}$  & &$              $ & 1784  & 1810  &  &       &  -0.14  &  -0.31  &  0.94   \\
 $\Xi''_{1/2}$     & &$              $ & 1922  & 1930  &  &       &  -0.14  &  -0.31  &  0.94   \\
 $\Omega_{1/2}$  & &$              $ & 2061  & 2020  &  &       &        &        &  1.00   \\
\hline
 $\Delta_{3/2}$  & &$ 1720 \pm 50  $ & 1720  & 1685  &  &       &        &        &  1.00   \\
 $\Sigma''_{3/2}$  & &$              $ & 1847  & 1805  &  &       &  -0.19  &  (-0.80)  &  (0.57)   \\
 $\Xi''_{3/2}$     & &$              $ & 1973  & 1920  &  &       &  -0.19  &  (-0.80)  &  (0.57)   \\
 $\Omega_{3/2}$    & &$              $ & 2100  & 2020  &  &       &             &             &  1.00   \\
\hline
 $\Lambda''_{1/2}$ & &$ 1407 \pm 4   $ & 1407  & 1490  &  & 0.97  & 0.23  & 0.04  &         \\
\hline
 $\Lambda''_{3/2}$ & &$ 1520 \pm 1   $ & 1520  & 1490  &  & 0.98  & 0.18  & -0.01  &         \\
\hline \hline
\end{tabular}
\end{center}
\end{table}


\begin{table}[ht]
\caption{ Partial contributions to masses (in MeV) by singlet operators.
The first two rows contain the off diagonal contributions.}
\label{tab7}
\begin{center}
\begin{tabular}{|c|c|c|c|c|c|c|c|c|c|c|c|c|}
\hline
 & $c_1 O_1$ & $c_2 O_2$ & $c_3 O_3$ & $c_4 O_4$ & $c_5 O_5$ & $c_6 O_6$ &$c_7 O_7$
 & $c_8 O_8$ & $c_9 O_9$ & $c_{10} O_{10}$ & $c_{11} O_{11}$ &  $O_{total}$ \\
\hline
$  N_{1/2} - N'_{1/2} $ &    0 & -17 & -57 &  12 &   8 &   0 &   0 & 1 &   5 &   0 &  -8 &  -56\\
$  N_{3/2} - N'_{3/2} $ &    0 & -27 &  18 &  19 &  13 &   0 &   0 & 0 &   7 &   0 & -13 &   17\\
$ N_{1/2}  $ & 1347 & -17 &   0 &  86 & -16 & 160 &  27 & 0 & -25 &  16 &   4 & 1580\\
$ N_{3/2}  $ & 1347 &   9 &   0 & -43 &   8 & 160 &  27 & 0 &  13 &  16 &  -2 & 1534\\
$ N'_{1/2} $ & 1347 & -43 & -40 &  76 & -41 & 320 & -27 & 2 &  16 &  31 & -21 & 1621\\
$ N'_{3/2} $ & 1347 & -17 &  32 &  31 & -16 & 320 & -27 &-1 &   7 &  31 &  -8 & 1698\\
$ N_{5/2}  $ & 1347 &  26 &  -8 & -46 &  25 & 320 & -27 & 0 & -10 &  31 &  12 & 1671\\
$\Delta_{1/2}  $ & 1347 &  17 &   0 & -49 & -33 & 320 &  53 & 0 &  21 & -25 &  -7 & 1645\\
$\Delta_{3/2}  $ & 1347 &  -9 &   0 &  24 &  16 & 320 &  53 & 0 & -11 & -25 &   3 & 1720\\
$\Lambda''_{1/2} $ & 1347 & -52 &   0 &   0 &   0 &   0 &   0 & 0 &   0 &   0 &   0 & 1295\\
$\Lambda''_{3/2} $ & 1347 &  26 &   0 &   0 &   0 &   0 &   0 & 0 &   0 &   0 &   0 & 1373\\
\hline
\end{tabular}
\end{center}
\end{table}


\begin{table}[ht]
\caption{ Partial fits of non-strange masses and the singlet $\Lambda$ masses using subsets of singlet operators. The mixing angles are predicted except for the last row. The last row results from solving for the coefficients $c_1,\cdots,c_{11}$ with the eleven inputs given by the seven  non-strange masses, the two  singlet $\Lambda$ masses and the two leading order mixing angles. The singlet $\Lambda$ masses are corrected by a shift of $135~{\rm MeV}$ to include
$SU(3)$ breaking. }
\label{tab8}
\begin{center}
\begin{tabular}{|c|c|c|c|c|c|c|c|c|c|c|c|c|c|c|}
\hline
$c_1$ & $c_2$ & $c_3$ & $c_4$ & $c_5$ & $c_6$ & $c_7$ & $c_8$ & $c_9$ & $c_{10}$ & $c_{11}$ &$\chi^2/dof$ & $\theta_1$ & $\theta_2$ \\
\hline
 517  &  -    &  -    &  -    & -     &  -    & -     & -     & -     & -        & -        & 99.1      & 0    & 0.  \\
 448  &  -    &  -    &  -    & -     & 508   & -     & -     & -     & -        & -        &  5.4      & 0    & 0.  \\
 447  &  -    &  40   &  -    & -     & 514   & -     & -     & -     & -        & -        &  6.1      & 0.12 & 3.11 \\
 453  & 64    &  -    & 51    & -     & 474   & -     & -     & -     & -        & -        &  2.2      & 0.13 & 0.15 \\
 454  & 69    &  68   & 63    & -     & 481   & -     & -     & -     & -        & -        &  2.3      & 0.40 & 0.08 \\
 451  & 76    & 113   & 114   & 79    & 507   & -122  & -     & -     & -        & -        &  1.1     & 0.72 & 3.08 \\
\hline
 449  & 75    & 123   & 109   & 77    & 487   & -125  & 9    & 16    & -64     & 18       &  exact   & 0.61 & 3.04 \\
\hline
\end{tabular}
\end{center}
\end{table}

\pagebreak

\begin{thebibliography}{99}

\bibitem{CandR}
S.~Capstick and W.~Roberts,
Prog.~Part.~Nucl.~Phys.~{\bf 45}, S241 (2000).

\bibitem{ChPT}  U.-G.~Mei\ss ner, in \lq\lq Encyclopedia of Analytic QCD\rq\rq,  Vol. 1, 417 (2000), M.~Shifman editor; World Scientific (2000).

\bibitem{RevMan}
A.~V.~Manohar,
\lq\lq Large N QCD\rq\rq, hep-ph/9802419.
Published in Proceedings of \lq\lq Probing the Standard Model of Particle Interactions\rq\rq.~F.~David and R.~Gupta editors.

\bibitem{RevJen}
E.~Jenkins,
Ann.~Rev.~Nucl.~and~Part.~Sci.~{\bf 48},  81 (1998).

\bibitem{RevLeb}
R.~F.~Lebed,
Czech.~J.~Phys.~{\bf 49},  1273 (1999).~

\bibitem{Flores}
P.~Bedaque and M.~Luty,
Phys.~Rev. {\bf D54}, 2317 (1996).\\
Y.~Oh and W.~Weise,
Eur.\ Phys.\ J.\ A {\bf 4}, 363 (1999).\\
R.~Flores-Mendieta, Ch.~P.~Hofmann, E.~Jenkins and  A.~V.~Manohar,
Phys.~Rev.~{\bf D62}, 034001 (2000).

\bibitem{CGO94}
C.~D.~Carone, H.~Georgi, L.~Kaplan and D.~Morin,
Phys.~Rev. {\bf D50}, 5793 (1994).

\bibitem{Goi97}
J.~L.~Goity,
Phys.~Lett.~{\bf B414}, 140 (1997).

\bibitem{PY98}
D.~Pirjol and T.-M.~Yan,
Phys.~Rev. {\bf D57}, 1449 (1998);
{\bf D57},  5434(1998).

\bibitem{CCGL}
C.~E.~Carlson, C.~D.~Carone, J.~L.~Goity, and R.~F.~Lebed,
Phys.~Lett.~{\bf B438}, 327 (1998);
Phys.~Rev.~{\bf D59}, 114008  (1999).

\bibitem{SGS02}
C.~L.~Schat, J.~L.~Goity and N.~N.~Scoccola,
Phys.~Rev.~Lett.~{\bf 88}, 102002 (2002).

\bibitem{Schat} C.~L.~Schat, hep-ph/0204044. 

\bibitem{Azi01} Z.~A.~Baccouche, C.~K.~Chow, T.~D.~Cohen and B.~A.~Gelman, Nucl.~Phys. {\bf A696}, 638 (2001).

\bibitem{CLAS}
V.~Burkert,
\lq\lq The $N^*$ program at Jefferson Lab: status and prospects\rq\rq, hep-ph/0207149 and references therein.

\bibitem{Ric01}
D.~G.~Richards {\it et al.}, (QCDSF/UKQCD/LHPC Coll.),
Nucl.~Phys.~Proc.~Suppl.~{\bf 109}, 89 (2002).\\
M.~G\"ockeler {\it et al.},
(QCDSF/UKQCD/LHPC Collaboration), Phys.~Lett.~{\bf B532}, 63 (2002).\\
W.~Melnitchouk {\it et al.}, hep-lat/0202022, and
Nucl.~Phys.~Proc.~Suppl.~{\bf 109}, 96 (2002).\\
J.~M.~Zanotti {\it et al.}  (CSSM Lattice Collaboration),
Phys.~Rev.~{\bf D65}, 074507 (2002).\\
S.~Sasaki, T.~Blum and S.~Ohta,
Phys.~Rev.~{\bf D65}, 074503 (2002).

\bibitem{Simonov}
Yu.~A.~Simonov, Phys.~Rev.~{\bf D65}, 116004 (2002).\\
Yu.~A.~Simonov,  hep-ph/0205334.\\
F.~X.~Lee and X.~Liu, Phys.~Rev.~{\bf D66}, 014014 (2002).

\bibitem{tHo74}
G.~'t Hooft, Nucl.\ Phys.\ {\bf B72}, 461 (1974).

\bibitem{Witten}
E.~Witten, Nucl.~Phys  {\bf B160}, 57 (1979).~

\bibitem{GeSa84}
J.~L.~Gervais and B.~Sakita,
Phys.~Rev.~Lett. {\bf 52}, 87 (1984);
Phys.~Rev.  {\bf D30}, 1795 (1984).\\
K.~Bardakci,
Nucl.~Phys. {\bf B243} 197 (1984).

\bibitem{Dashen1}
R.~Dashen and A.~V.~Manohar,
Phys.~Lett. {\bf B315}, 425 (1993);
Phys.~Lett. {\bf B315}, 438 (1993).

\bibitem{LM94}
C.~D.~Carone, H.~Georgi and S.~Osofsky,
Phys.~Lett. {\bf B322}, 227 (1994).\\
M.~A.~Luty and J.~March-Russell,
Nucl.~Phys. {\bf B426}, 71 (1994).\\
M.~A.~Luty, J.~March-Russell and  M.~White,
Phys.~Rev. {\bf D51}, 2332 (1995).


\bibitem{Jenk1}
E.~Jenkins,
Phys.~Lett.  {\bf B315}, 441 (1993).

\bibitem{Jenk2}
R.~Dashen, E.~Jenkins, and A.~V.~Manohar,
Phys.~Rev. {\bf D49}, 4713 (1994).

\bibitem{Jenk3} E.~Jenkins and A.~V.~Manohar,
Phys.~Lett. {\bf B335},  452 (1994).

\bibitem{DJM2}
R.~Dashen, E.~Jenkins, and A.~V.~Manohar,
Phys.~Rev. {\bf D51}, 3697 (1995).




\bibitem{JL95}
E.~Jenkins and R.~F.~Lebed,
Phys.~Rev. {\bf D52}, 282 (1995).

\bibitem{Greenberg67}
O.~W.~Greenberg and M.~Resnikoff,
Phys.~Rev. {\bf 163}, 1844 (1967).


\bibitem{Dai}
J.~Dai, R.~Dashen, E.~Jenkins, and A.~V.~Manohar,
Phys.~Rev.  {\bf D53}, 273 (1996).

\bibitem{Buchmann} 
A.~J.~Buchmann, J.~A.~Hester and R.~F.~Lebed, hep-ph/0205108.\\
A.~J.~Buchmann and R.~F.~Lebed, hep-ph/0207358.\\
E.~Jenkins, X.~Ji and A.~V.~Manohar, hep-ph/0207092.


\bibitem{GS}
J.~L.~Goity and C.~L.~Schat,
in preparation.


\bibitem{IK1}
N.~Isgur and G.~Karl, Phys.~Rev. {\bf D18}, 4187 (1978).

\bibitem{IC} S.~Capstick and N.~Isgur, Phys.\ Rev.~{\bf D34}, 2809 (1986).


\bibitem{CC00}
C.~E.~Carlson and C.~D.~Carone,
Phys.~Lett. {\bf B441}, 363 (1998); Phys.~Rev. {\bf D58}, 053005 (1998); Phys.~Lett. {\bf  B484}, 260 (2000).


\bibitem{Mattis}
M.~P.~Mattis and M.~Karliner,
Phys.~Rev.~{\bf D31}, 2833 (1985); {\it ibid.} {\bf D34}, 1991 (1986).\\
G.~Holzwarth and B.~Schwesinger,
Rep.~Prog.~Phys. {\bf 49}, 825  (1986).\\
I.~Zahed and G.~E.~Brown,
Phys.~Rep. {\bf 142},  1 (1986).\\
B.~Schwesinger, H.~Weigel, G.~Holzwarth and A.~Hayashi,
Phys.~Rep. {\bf 173 } (1989) 173.

\bibitem{Lee02}
E.~Oset, A.~Ramos and C.~Bennhold,
Phys.~Lett.  {\bf B527}, 99 (2002).\\
M.~F.~Lutz and E.~E.~Kolomeitsev,
Nucl.~Phys. {\bf A700}, 193 (2002).

\bibitem{PDG}
Particle Data Group (D.~E.~Groom {\it et al.}),
Eur.~Phys.~J. {\bf C15}, 1 (2000).

\bibitem{Edm57}
A.~R.~Edmonds,
Angular Momentum in Quatum Mechanics (Priceton Univ.~Press, New Jersey, 1974).

\bibitem{Hec62}
K.~T.~Hecht,
Nucl.~Phys. {\bf 62}, 1 (1965).
\bibitem{Ver68}
J.~D.~Vergados,
Nucl.~Phys. {\bf A111}, 681 (1968).

\bibitem{footone} Interestingly, in the case of the S representation the matrix elements of
the excited quark spin are of order $1/N_c$, and therefore in that case
spin-orbit interactions will be of order $1/N_c$.

\bibitem{Glozman}
L.~Ya.~Glozman and  D.~O.~Riska,
Phys.~Rept.~{\bf 268}, 263 (1996).

\bibitem{Collins}
H.~Collins and H.~Georgi,
Phys.~Rev. {\bf D59}, 094010 (1999).

\bibitem{AD73}
Y.~Akiyama and J.~P.~Draayer,
Comput.~Phys.~Commun.~{\bf 5}, 405 (1973).

\bibitem{HLC75}
A.~J.~Hey, P.~J.~Litchfield and R.~J.~Cashmore,
Nucl.~Phys.~{\bf B95}, 516 (1975).\\
D.~Faiman and D.~E.~Plane,
 Nucl.~Phys.~{\bf B50}, 379 (1972).

\bibitem{DGG75}
A.~De R\'ujula, H.~Georgi and S.~L.~Glashow,
Phys.~Rev. {\bf D12}, 147 (1975).



\bibitem{Schnitzer}
H.~Schnitzer,
Phys.~Lett.~{\bf B76}, 461 (1978).

\bibitem{Ginocchio}
J.~N.~Ginocchio,
Phys.~Rev.~Lett.~{\bf 78}, 436 (1997).\\
P.~R.~Page, T.~Goldman and J.~N.~Ginocchio,
Phys.~Rev.~Lett.~{\bf 86}, 204 (2001).


\end{thebibliography}
\end{document}